\documentclass[
preprint,
prb,aps,amsmath,amssymb,showpacs,floatfix
]
{revtex4-1}

\usepackage{graphicx}
\usepackage{dcolumn}
\usepackage{bm}
\usepackage{amsfonts}
\usepackage{color}
\usepackage{times}

\begin{document}

\title{Modulation of electron carrier density at the $n$-type  
LaAlO$_3$/SrTiO$_3$ interface by water adsorption}

\author{Yun Li}
\author{Jaejun Yu}
\email[Corresponding author.\ ]{Email: jyu@snu.ac.kr}
\affiliation{Department of Physics and Astronomy, Center for Strongly
Correlated Materials Research, Seoul National University, Seoul 151-747,
Korea}

\begin{abstract}
We investigate energetic stability and dissociation dynamics of  water
adsorption at the LaAlO$_3$ surface of the 
$n$-type  LaAlO$_3$/SrTiO$_3$ (LAO/STO) interface and its effect on  
electronic properties of the interface by carrying out first-principles 
electronic structure calculations. In an ambient 
atmosphere at room temperature  the configuration of 1 monolayer
(ML) of water molecules including 3/4 ML of dissociated water 
molecules adsorbed at the surface is found to be most stable, the
configuration of 1 ML of dissociated water molecules is metastable. 
Water molecule dissociation induces a shift-up of the valence band maximum
(VBM) of  the LAO surface, reducing the gap between the VBM of the LAO
surface and the conduction band minimum of the STO. For the LAO/STO interface
with three LAO unit-cell layers, once the coverage of
dissociated water molecules reaches 1/2 ML the gap is closed, the
interface becomes metallic and the carrier density at the LAO/STO
interface increases with increasing the coverage of dissociated water molecules.
Our findings suggest two ways to control the conductivity at the LAO/STO
interface: (I) insulator-metal transition by adsorbing a mount of water at the
bare surface; (II) carrier density change by the transition between the
most stable and the metastable adsorption configurations for 1 ML coverage in
an ambient atmosphere at room temperature. 
\end{abstract}

\maketitle

\section{Introduction}
Two-dimensional electron 
gas (2DEG) at interfaces has been one of the most intriguing 
subjects of electronic material research. Because of its 
versatile and remarkable properties many attentions have been 
drawn from both fundamental researches and application studies. 
One of the key findings in this subject is that a 2DEG can be 
generated at the interface between two insulating metal oxides, 
LaAlO$_3$/SrTiO$_3$ (LAO/STO) \cite{ohtomo2004}. At the $n$-type 
LAO/STO interface between (LaO)$^{+}$ and (TiO$_2$)$^0$ layers, a 
high density (a few 10$^{13}$ cm$^{-2}$) and  high mobility 
($\sim$ 10$^4$ cm$^2$V$^{-1}$s$^{-1}$) electron gas has been 
realized at low temperature\cite{ohtomo2004}. In addition, 
ferromagnetism\cite{brinkman2007}  and 
superconductivity\cite{reyren2007} were also reported at the same 
interface. As LAO and STO are two band insulators, finding the 
electron gas at their interface immediately triggered extensive 
researches to reveal the formation mechanism of the electron gas. 
Up to now, two mechanisms have been proposed and well accepted. 
One is the polar catastrophe, in which the 
polar electric field in the LAO film shifts up the valence band 
maximum (VBM) of the LAO film layer by layer and finally leads to a charge 
transfer from the valence band of the LAO surface layer to the 
conduction band of the STO\cite{nakagawa2006}. 
It was found in Thiel \textit{et al.}'s experiment 
and confirmed theoretically that there is a 
critical thickness of the LAO film (4 unit cell) below which the interface is 
insulting and above which the interface becomes 
metallic\cite{thiel2006,pentcheva2009,li2010}.  
The other mechanism is the role of 
oxygen vacancies, which were explicitly demonstrated in experiments to exist at 
the STO side for the samples grown at low
oxygen pressures\cite{herranz2007,kalabukhov2007}.  Theoretical study
showed that oxygen vacancies also readily form at the LAO surface and  produce a
carrier charge with a density of no more than half electron per u.c. area at
the interface\cite{li2011}. 

Recently, Cen \textit{et al.} reported a novel method to generate an electron  
gas at the LAO/STO interface\cite{cen2008,cen2009}. In their experiment,  
 nanoscale control of so-called insulator-to-metal transition (IMT)  
at the LAO/STO interface was realized by an atomic-force-microscopy  
(AFM) tip scanning the LAO surface. A nanoscale  
conducting line can be formed at the interface by the AFM tip with  
a positive bias voltage  \textit{writing} on the LAO surface and  
the conducting line can be erased by the AFM tip with a negative  
bias voltage \textit{cutting} the line. It is noticeable that this  
IMT only happens to the interface  
with three unit-cell (u.c.) layers of LAO. Subsequent experimental studies  
suggested that this IMT may be related to water adsorption at the  
LAO surface\cite{bi2010,xie2011}. To interpret this phenomenon several  
assumptions have been proposed. Bi \textit{et al.} assumed that  
the adsorbed water are dissociative and the AFM tip with a  
positive bias voltage removes the hydroxyls (OH$^{-}$) from the  
surface, as a result the left hydrogen atoms dope electron carriers at  
the interface\cite{bi2010}. However, first-principles electronic  
structure calculations have showed that regardless of the LAO  
thickness hydrogen adsorption on the LAO  
surface always induces a metallic interface\cite{son2010}, implying that the  
change at the LAO surface induced by the AFM tip scanning is not like  
Bi \textit{et al.}'s scenario. Xie \textit{et al.} speculated  
that water molecules are adsorbed at the LAO surface by a  
dipole-dipole interaction (the dipole of water molecule and dipole of the LAO  
film)\cite{xie2011}, i.e., two hydrogen atoms of a water molecule approach to  
the LAO surface  
and the oxygen atom points to the outside. This will lead to an  
enhanced polarity of LAO film with adsorbed water compared to  
the bare LAO film, probably resulting in a charge transfer from  
the surface to the interface. However, many studies  
have shown that water adsorption on oxide surfaces is via the  
formation of a chemical bond between the oxygen atom of the water 
molecule and the metal atom at the oxide 
surfaces\cite{thiel1987,hass1998,shapovalov2000,henderson2002,ionescu2002,
 evarestov2007, ranea2008,baniecki2009,guhl2010} instead of a dipole-dipole
interaction, basically ruling out Xie 
\textit{et al.}'s scenario. This is because  
that the energy gain of the chemical adsorption on oxide surfaces, 
1$\sim$3 eV per water 
 molecule\cite{hass1998,evarestov2007,ranea2008,guhl2010},  is far 
larger than that of the physical adsorption, about 0.2 eV per  
molecule\cite{emsley1980}. 
Up to now, although this nanoscale  
control of transport property at the LAO/STO interface has presented a  
very promising application on nanoscale memory device, called  
'atomic force pencil and eraser'\cite{rijnders2008},  its physical  
nature is still a puzzle.

In this paper we present a first-principles study of
electronic and structural characteristics of 
water molecules adsorbed at the LAO surface of the $n$-type
LAO/STO interface, including surface atomic structures, thermodynamic 
stabilities, dissociation processes of water molecules, and electronic 
properties. We find that in an ambient atmosphere  at room temperature the
adsorption configuration of one monolayer (ML) of water molecules including
3/4 ML of dissociated water molecules is most stable and the configuration of
1 ML of dissociated water molecules  is metastable. The dissociation of 
water molecules at the LAO surface can reduce and even close the gap between
the VBM of the LAO surface and the conduction band minimum (CBM) of the STO,
leading to a metallic interface. Remarkably, for the ground-state and metastable
adsorption  configurations electron carrier densities at the interface show a
large  difference. Based on our findings we make an attempt to reveal the
mechanism of the mysterious IMT found by  Cen \textit{et al.}.

\section{Computational details}
We carried out first-principles electronic structure calculations
within the frame of density functional theory as implemented in
the Vienna \textit{ab initio} simulation package
(VASP)\cite{kresse1996}. A generalized gradient approximation
(GGA)\cite{wang1991} for exchange-correlation functional and the
projector augmented wave method\cite{blochl1994,kresse1999} with a
kinetic cutoff energy of 400 eV for the plane wave basis set were
used. The $n$-type LAO/STO interface was modelled into a
(2$\times$2) supercell consisting of four STO(001) layers, three
LAO layers, and a vacuum region of about 14 {\AA}. Various
coverages of water molecules were located at the
AlO$_2$-terminated surface. Dipole correction was employed to
correct the errors of electrostatic potential, atomic force and
total energy caused by periodic boundary
condition\cite{makov1995}. The in-plane lattice constant of the
slab was constrained at the calculated equilibrium value of STO
bulk ($a=3.94$ {\AA}). A (4$\times$4$\times$1)  \textbf{k}-point
grid was used for calculations of atomic structure
optimization  and a (6$\times$6$\times$1) \textbf{k}-point grid
for electronic structure calculations. All atomic
coordinates were fully relaxed until the atomic forces were less
than 0.02 eV/{\AA}, except for the atoms in bottom two STO layers,
which were fixed in their bulk positions. A first-principles
molecular dynamics simulation at T=300 K was performed to search
the atomic structure of 2 ML of water molecules adsorbed at the
surface. The nudged elastic band method was used to calculate
reaction paths of water molecule dissociation at the surface.

\section{Thermodynamic stability of water adsorption}

\begin{figure*}[htbp]
\centering
\includegraphics[width=0.9\textwidth]{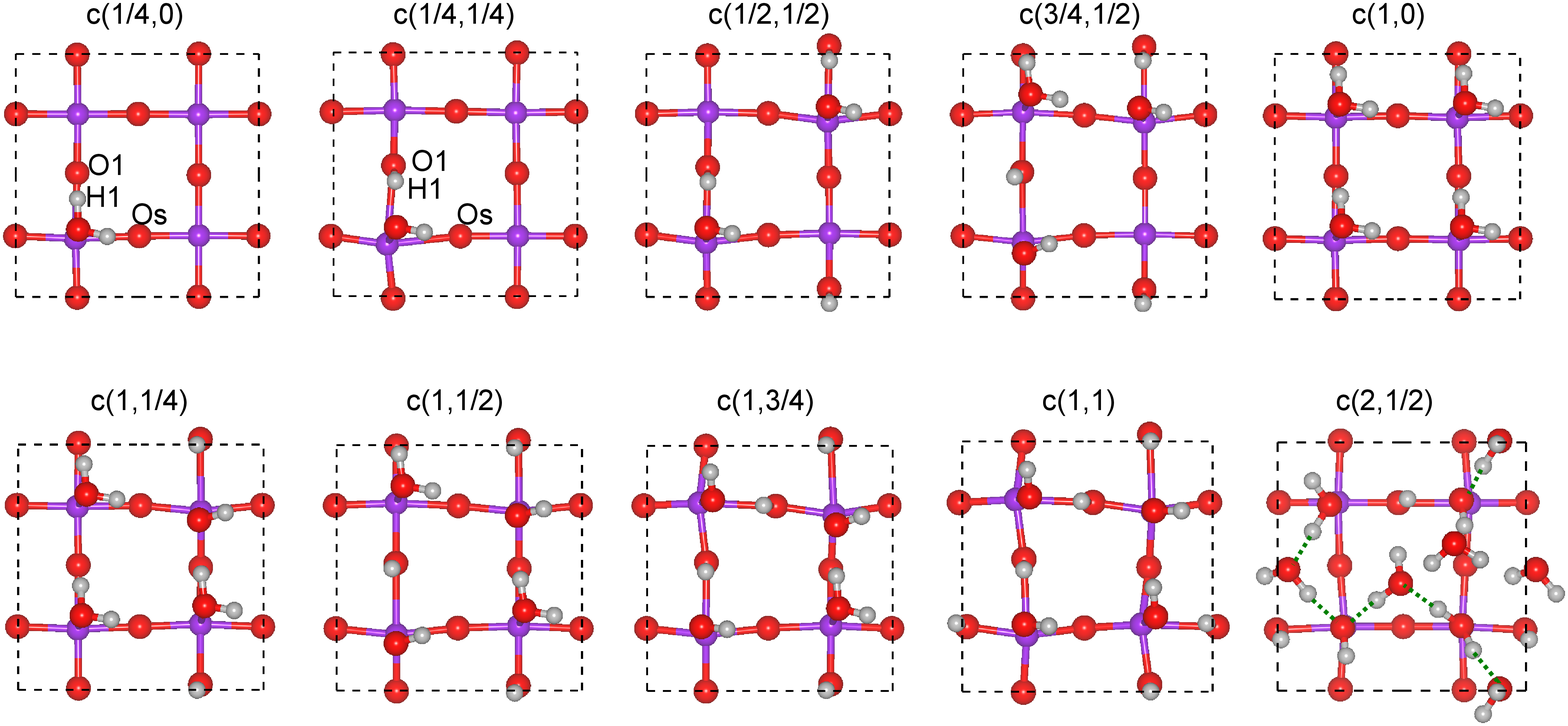}
\caption{\label{fig:fig1}(Color online) Top views of
adsorption configurations of water molecules adsorbed 
at the AlO$_2$-terminated surface
of the n-type LAO/STO interface. The notation
$c(\theta,\delta)$ denotes the adsorption configurations, in
which $\theta$ is the total coverage of water molecules and
$\delta$ is the coverage of dissociated molecules in unit of ML.
Purple balls represent Al atoms, red O atoms, and grey H atoms. In
$c(2,1/2)$ green dot lines represent hydrogen bonds.}
\end{figure*}

Figure~\ref{fig:fig1} displays geometries of various adsorption configurations
at different coverages, in which water molecules are either
molecular or dissociated. For the molecular adsorption, for instance
$c(1/4,0)$, the O atom of the water molecule sits on the Al atom
forming an Al-O bond, and two H atoms approach to two closely neighbouring
O atoms at the AlO$_2$ layer. For the dissociative adsorption, for instance
$c(1/4,1/4)$, two distinct hydroxyl groups are produced, one of
which originates from the water molecule and sits on the Al atom,
the other (O1-H1) lies in the AlO$_2$ layer. For
the configurations of no more than 1 ML coverage water molecules
are bound to the surface by valence bonds or ionic bonds. Once
the coverage exceeds 1 ML, as shown in $c(2,1/2)$, the extra
molecules are loosely adsorbed at the surface by hydrogen
bonds.

\begin{figure}[htbp]
\centering
\includegraphics[width=0.4\textwidth]{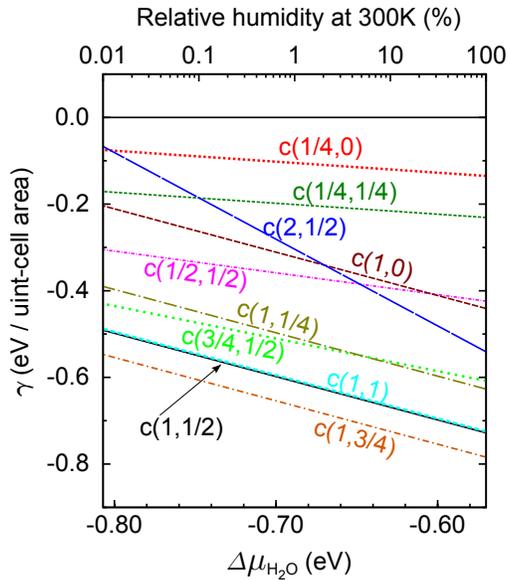}
\caption{\label{fig:fig2} (Color online) Surface free energies of
various adsorption configurations at the LAO surface as functions of the
relative chemical potential of water vapor,
$\Delta\mu_{\rm H_2O}$.
The top axis shows the relative humidity of the ambient atmosphere at 300 K
corresponding to the relative chemical potential on the bottom axis. The free
energy of the bare surface is taken to be zero.} 
\end{figure}

Since the IMT observed in Cen \textit{et al.}'s experiments happened in a
humid ambient atmosphere at room temperature, to directly address the
experimental results we first investigated thermodynamic stabilities of
various adsorption configurations in an ambient atmosphere at room
temperature by comparing their surface free energies. We calculated the surface
free energies by employing the first-principles thermodynamics
approach\cite{guhl2010}, in which the surface is assumed to be in equilibrium
with a humid ambient atmosphere characterized by the
relative chemical potential of  water vapor. 
The most stable configuration at a given relative chemical potential minimizes
the surface free energy approximately defined as
$$
 \gamma=\frac{1}{A}\{ E_{\rm ad} - E_{\rm bare} -
n_{\rm H_2O} \cdot [ E_{\rm H_2O(gas)}
 + \Delta\mu_{\rm H_2O}(T,p_{\rm H_2O})] \}
\ \ . $$ 
Here, $E_{\rm ad}$  is the total energy of the interface
with $n_{\rm H_2O}$  adsorbed water molecules per
surface area $A$, $E_{\rm bare}$ is the total energy of the
bare surface, $E_{\rm H_2O(gas)}$ is the total energy of a gaseous
water molecule, 
 $\Delta\mu_{\rm H_2O}(T,p_{\rm H_2O})$
is the relative chemical potential of the water vapor at the
temperature $T$ and partial pressure $p_{\rm H_2O}$. 
The vibrational free energy and
configuration entropy contributions are neglected in the
definition of surface free energy.
$E_{\rm ad}$, $E_{\rm bare}$, and $E_{\rm H_2O(gas)}$  were
obtained from our first-principles calculations.
$\Delta\mu_{\rm H_2O}(T,p_{\rm H_2O})$ follows from the
ideal-gas relation
$$
\Delta\mu_{\rm H_2O}(T,p_{\rm H_2O})=
\Delta\mu_{\rm H_2O}(T,p^{\circ}_{\rm H_2O}) +
k_{\rm B}T
ln(\frac{p_{\rm H_2O}}{p^{\circ}_{\rm H_2O}}) \ .
$$
At standard pressure $p^{\circ}_{\rm H_2O}$=1 bar  the
relative chemical potential can be derived using
$$
\Delta\mu_{\rm H_2O}(T,p^{\circ}_{\rm H_2O})
=[H(T,p^{\circ}_{\rm H_2O}) 
-H({\rm 0K}, p^{\circ}_{\rm H_2O})]
-T[S(T,p^{\circ}_{\rm H_2O})
-S({\rm 0K},p^{\circ}_{\rm H_2O})] \ ,
$$
where the enthalpy and entropy differences are obtained from
thermochemical tables\cite{stull1971}.
The ambient atmosphere at
room temperature sets an upper limit for
$\Delta\mu_{\rm H_2O}=-0.57$ eV per water molecule,
which corresponds to the relative chemical potential of water molecules in a
water-saturated
atmosphere (100\% relative humidity)  at room temperature. The chemical
potential at any other relative humidity can be obtained from the above
ideal-gas relation. Figure~\ref{fig:fig2} displays
the surface free energies of various adsorption configurations
varying with the relative chemical potential or corresponding relative humidity
at room temperature.
In the range of relative humidity varying from 0.01\% to 100\%, which
nearly covers all possible relative humidities of the ambient atmosphere, 
the most stable configuration is $c(1,3/4)$. The configurations
with a lower water coverage, such as $c(1/4,1/4)$, can only exist in an
extremely dry environment. Noticeably, the configuration
$c(2,1/2)$ with 2 ML of water has a higher free energy than $c(1,3/4)$ and even
than $c(1,1/2)$. This implies that once the coverage reaches 1 ML the
surface will not adsorb water molecules from the atmosphere
anymore, instead, water molecules readily remain in the
atmosphere.

\section{Adsorption of 1/4 ML of water molecules: dissociation process and
electronic properties}

\begin{figure}[htbp]
\centering
\includegraphics[width=0.5\textwidth]{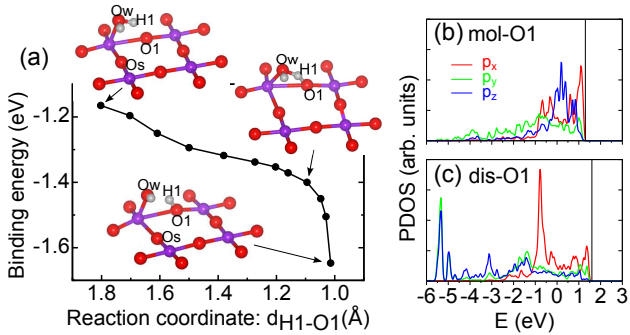}
\caption{\label{fig:fig3}(Color online) (a) Dissociation process of
1/4 ML of water molecules adsorbed at the AlO$_2$-terminated surface of the
LAO/STO interface. Reaction coordinate is the distance
d$_{\rm O1-H1}$ between the hydrogen atom H1 and the oxygen
atom O1. Insets are perspective views for molecular, intermediate and
dissociative configurations. (b), (c) Projected densities of
states (PDOSs) for O1 atom in (b) the molecular (mol) and (c) dissociative
(dis) adsorption configurations of 1/4 ML coverage, respectively. 
The vertical lines denote the Fermi level. The VBM of the STO bottom
layer is taken to zero (see Fig.~\ref{fig:fig4}). The $x$ and $y$ axes
are approximately along the direction of Al-Os-Al chain and that of Al-O1-Al
chain shown in the insets in (a).}
\end{figure}

For simplicity and clarity we first explored the dissociation process of 1/4 ML
of water molecules adsorbed at the LAO surface and its effect on electronic
properties of the interface system.  
Figure~\ref{fig:fig3}(a) displays a dissociation process without
energetic barrier for 1/4 ML of water molecules at the LAO
surface, in which the binding energy is defined as
$E_{\rm b}=E_{\rm ad}-E_{\rm bare}-n_{\rm H_2O}\cdot
E_{\rm H_2O(gas)}$. The intermediate state involves
 a four-member ring Al-Ow-H1-O1. Generally, a water molecule
dissociation at metal oxide surfaces is associated with a breaking
of the metal-oxygen bond at the
surface\cite{hass1998,evarestov2007,ranea2008,guhl2010,shin2010},
implying that the dissociation occurs more easily at the surface
with loose metal-oxygen bonds. Indeed, the Al-O bond at the LAO
surface is far weaker than  that of $\alpha$-Al$_2$O$_3$. This
fact is reflected by the bond lengths in these two materials, 1.95
{\AA} (in-plane) for the former and 1.70 {\AA} for the
latter\cite{ranea2008}. This implies that the energy cost to
stretch the Al-O1 bond in the four-member ring is relatively low
and can be compensated by the formation of hydroxyl O1-H1 bond. To
reveal the driving force of water molecule dissociation at the LAO
surface we checked on the change of electronic states of O1 atom
in the molecular and dissociative adsorptions.
In the molecular adsorption [Fig.~\ref{fig:fig3}(b)] the $p_z$
orbital of O1 is a non-bonding state lying near the top of valence
band. While in the dissociative adsorption
[Fig.~\ref{fig:fig3}(c)] the bonding state of the hydroxyl O1-H1
mixing part states of $p_z$ and $p_y$ orbitals lies at the bottom
of valence band. The lower surface energy of the dissociative
adsorption just arises from the falling down of band energies of
these O $p$ states. Therefore, we see in Fig.~\ref{fig:fig2} that 
for the configuration of 1 ML coverage the surface free energy
decreases with increasing the coverage of dissociated molecules up
to 3/4 ML. The energy rising of 1 ML of dissociated  water molecules
is due to a large structural distortion at the surface.

\begin{figure}[htbp]
\centering
\includegraphics[width=0.45\textwidth]{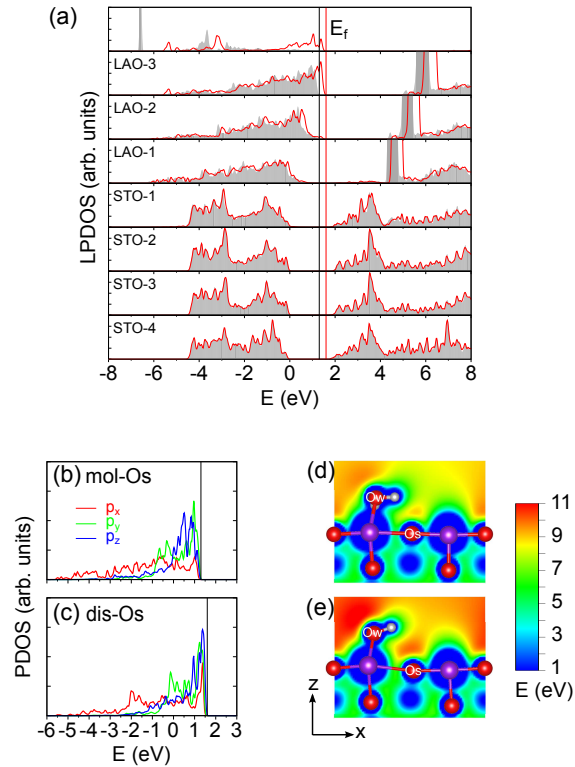}
\caption{\label{fig:fig4}(Color online)
(a) Layer-resolved projected densities
of states (LPDOSs) for the configurations of 1/4 ML coverage.
Grey-filled lines represent the LPDOSs
for the molecular adsorption, red solid lines represent the LPDOSs for
the  dissociative adsorption. The PDOSs of water molecule are amplified
four times for clear vision.
(b), (c) Projected densities of
states (PDOSs) for Os atoms in (b) the molecular (mol) and  (c) dissociative
(dis) adsorption configurations of 1/4 ML coverage, respectively.
The vertical lines denote the Fermi level. The VBM of the STO bottom
layer is taken to be zero.
(d), (e) Potential energies averaged along the $y$ axis over the slab of 2
{\AA} centering at Os atom in (d) the molecular and (e)
dissociative adsorption configurations of 1/4 ML coverage. }
\end{figure}

Figure~\ref{fig:fig4}(a) shows LPDOSs of molecular and dissociative
adsorptions
of 1/4 ML coverage. We note that water molecule dissociation gives
rise to
a remarkable change of electronic states of the surface layer,
compared to the molecular adsorption the VBM of the LAO
surface layer shift up about 0.2 eV in the dissociative adsorption. PDOSs in
Figs.~\ref{fig:fig4}(b) and \ref{fig:fig4}(c)
show that this shift-up is
mainly due to the shift-up of the $p_z$ orbital of the O atoms like Os atom
near the hydroxyl Ow-H. 
We ascribe this shift-up to an extra repulsive Coulomb potential generated by 
the negative-charged hydroxyl Ow-H.  As shown in
Fig.~\ref{fig:fig4}(d) and \ref{fig:fig4}(e), compared
to the molecular adsorption, in the dissociative adsorption the potential energy
over Os atom  rises up significantly, implying that the $p_z$ orbital of Os atom
lies at a higher position. Moreover, our calculations show
that as more water molecules dissociate the VBM
of the LAO surface layer shifts up more.
A coverage of
dissociated water molecules as high as no less than 1/2 ML 
can even shift up the VBM of the LAO surface layer by about 0.6 eV, 
consequently closing the gap between the VBM of the LAO surface layer 
and the STO CBM and thus leading to a metallic interface.

\section{Adsorption of 1 ML of water molecules: dissociation process and
electronic properties}

\begin{figure*}[htbp]
\centering
\includegraphics[width=1\textwidth]{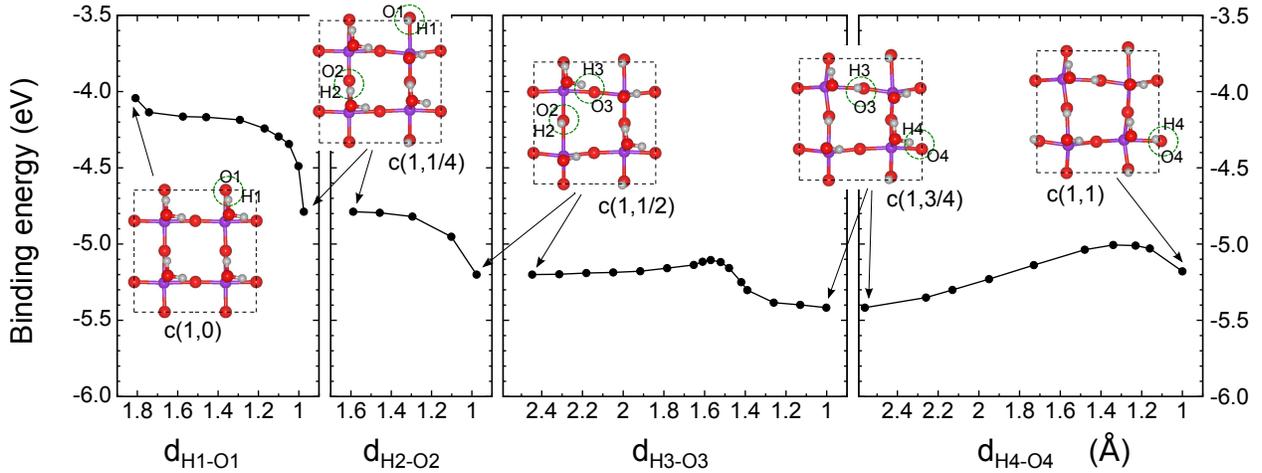}
\caption{\label{fig:fig5}(Color online) Dissociation process of 1 ML
of water molecules at the LAO surface.
Reaction coordinates are the distances d$_{\rm Hi-Oi}$ between the
hydrogen atom H$i$ and the oxygen atom O$i$, $i=1, 2, 3, 4$.}
\end{figure*}

Dissociation processes of 1/2, 3/4 and 1 ML of water molecules
adsorbed on the LAO surface were also investigated. We found that for all of
them there is an up-limit coverage of
water molecules that dissociate barrierlessly, 1/2 ML. Here we are
most interested in the case of 1 ML coverage because $c(1,3/4)$,
a configuration of 1 ML coverage, is the thermodynamic ground
state in an ambient atmosphere at room temperature and as shown
below the change of electronic properties of the LAO/STO interface
during the dissociation process of 1 ML of water molecules can be
compared with the experimental results. Figure~\ref{fig:fig5} shows the energy
profile of a reaction path for the  dissociation
of 1 ML of water molecules, in which four water molecules
dissociate one by one. We note that the barrierless dissociation
of preceding two water molecules induces a large distortion at the
surface. This distortion largely increases the distance between
the hydrogen atom H3 of the third water molecule and the oxygen
atom O3 at the LAO surface, finally giving rise to an energy
barrier of 0.1 eV in the dissociation path of the third water
molecule. Furthermore, the dissociation of the fourth water
molecule undergoes an even higher barrier of 0.41 eV and
the association of this dissociated molecule undergoes a barrier of 0.17 eV.

\begin{figure}[htbp]
\centering
\includegraphics[width=0.5\textwidth]{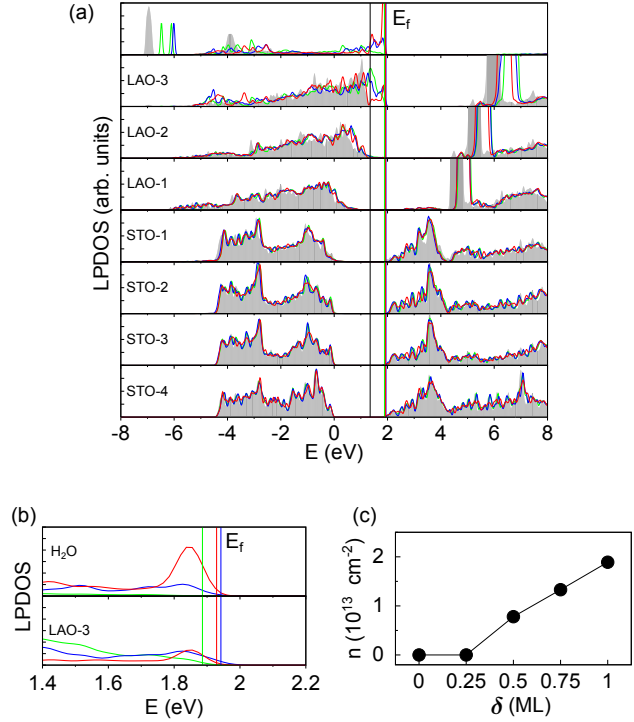}
\caption{\label{fig:fig6}(Color online) 
(a) Layer-resolved projected densities
of states (LPDOSs) for the configurations of 1 ML coverage.
Grey-filled lines represent the LPDOSs
for fully molecular adsorption $c(1,0)$ , green, blue, and red solid lines
represent the LPDOSs for $c(1,1/2)$, $c(1,3/4)$, and $c(1,1)$.
(b) Details of LPDOSs of water molecules and the LAO surface layer near the
Fermi level for $c(1,1/2)$, $c(1,3/4)$, and $c(1,1)$.
Lines are  colored as in (a).
Vertical solid lines denote Fermi levels.
The VBM of STO bottom layer is taken to 0 eV.
(c) Carrier densities for the configurations of 1 ML coverage with
different coverages of dissociated water molecules ($\delta$). }
\end{figure}

To clearly see the effect of the dissociation of 1 ML of water molecules  on
the electronic properties of the LAO/STO interface we plot
LPDOSs in Fig.~\ref{fig:fig6} (a) for 1 ML coverage.
PDOSs of water molecules and the LAO surface layer present apparent differences
in the molecular and dissociative adsorptions. 
In the fully molecular adsorption three
molecular orbitals of water molecule corresponding to three peaks ranging from
-7 to -2 eV are nearly well preserved. This molecular adsorption barely affects
the electronic states of the LAO surface, the band gap between the VBM of
the LAO surface and the CBM of the STO is 0.54 eV, approximately same as
the value of the interface with a bare surface, 0.56 eV. 
In contrast, water molecule dissociation significantly modifies 
the electronic states of the LAO surface and the interface. 
Accompanying with the dissociation the VBM of the LAO surface
layer shifts up considerably and a surface state originated from the
hydroxyls on Al atoms appears at the VBM. Strikingly, the dissociation of no
less than 0.5 ML of water molecules closes the band gap, 
resulting in a metallic interface.
Figure~\ref{fig:fig6}(b) shows the details of the electronic states of water
molecules and the LAO surface layer near the Fermi level. Since the VBM of the
LAO surface layer and the water molecules  exceeds the STO CBM, a part of
charge transfers from the surface into the STO conduction band. 
The carrier density is equal to the density of the empty states in the
valence band shown in Figure~\ref{fig:fig6}(b). 
As shown in  Figure~\ref{fig:fig6}(c), the 
carrier density increases with increasing the coverage of dissociated water
molecules. The carrier density is $1.3\times10^{13}$ cm$^{-2}$ for
$c(1,3/4)$ , and reaches to the highest value, $1.9\times10^{13}$ cm$^{-2}$, 
for $c(1,1)$.

\section{Discussions}
\begin{figure}[htbp]
\centering
\includegraphics[width=0.5\textwidth]{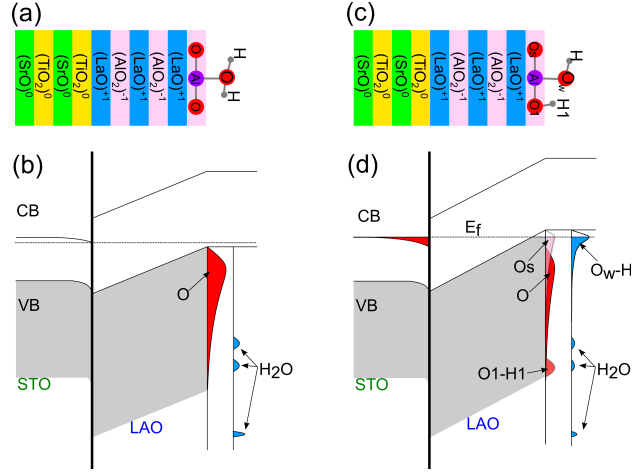}
\caption{\label{fig:fig7} (Color online) Schematic atomic structures
and band diagrams of the LAO/STO
interface with water adsorption on the surface. 
(a), (b)  Molecular adsorption,  and (c), (d) dissociative adsorption.
Shade areas are the valence bands of the STO and LAO. Red and pink filled
curves
present O $p$ states of the LAO surface layer, blue filled curves represent the
states of water molecules.}
\end{figure}

The above calculated results are schematically illustrated in
Fig.~\ref{fig:fig7}. The electronic properties at the LAO/STO
interface are nearly not affected by molecular adsorption of water
molecules at the LAO surface. Dissociative adsorption shifts up
the $p$ states of the O atoms (like Os) neighboring the
hydroxyl Ow-H on the Al atom, and induces a state originated from the
hydroxyl Ow-H at the VBM of the LAO surface. Once the
coverage of dissociated water molecules reaches to 1/2 ML the VBM
of the surface 
exceeds the STO CBM, resulting in a charge transfer from the surface to the
conduction band of the STO. In ambient atmosphere at room
temperature $c(1,3/4)$ is ground state and has a lower carrier
density, while $c(1,1)$ is metastable and has a higher carrier
density. This implies that a structural transition between these
two stable adsorption configurations can induce a conductance change at the
LAO/STO interface. This picture provides us a way to interpret the mechanism of
the IMT observed by Cen \textit{et
al.}. But, before elucidating it we should first clarify so-called
\textit{insulator} state in their experiment. 
In fact, before the AFM tip with
positive voltages scanning the surface the interface already
presents a conductance about 6.9 nS, which was referred to the
insulating state by  Cen \textit{et al.}, and after the AFM tip
scanning the surface the conductance rises to
about 7.6 nS. The ratio of conductance change is about ten percent.
In Comparison with about five orders of magnitude resistance
difference between the insulating and metallic interfaces reported
by Thiel \textit{et al.}\cite{thiel2006}, the interface with 6.9 nS
conductance should be rather regarded as metallic. Considering this point and
associating the conductance change observed in Cen \textit{et al.}'s experiment
with the difference of carrier densities of two stable adsorption
configurations, $c(1,3/4)$ and $c(1,1)$, we propose a mechanism  as the
following. Before the AFM tip scanning the LAO surface the configuration of
water molecules lies at the thermodynamic ground state $c(1,3/4)$, the
interface shows a relatively low conductance. After the AFM tip
with positive voltages scanning the LAO surface all water
molecules in the scanned area dissociate, i.e., the configuration
of water molecules in the scanned area becomes $c(1,1)$, thus the
interface shows a higher conductance.  
As for how the AFM tip dissociates water molecules, a recent study on water
molecule dissociation at a MgO surface by means of  STM tip scanning
has provided a possible answer\cite{shin2010}, in which it was demonstrated
that a tunnelling current passing
through the adsorbed water molecule can induce a lateral vibration
model of water molecule, consequently facilitating the 
dissociation. 
Therefore, it is highly possible that in Cen \textit{et al.}'s experiment 
water molecules in the
area scanned by the AFM tip were fully dissociated due to a tunnelling current
passing through them. 

Moreover, according to our calculations 
the carrier density change at the LAO/STO
interface induced by different water adsorptions
can only happen in the case of three LAO layers, being in good
agreement with Cen \textit{et al.}'s experimental result. 
For the interface with two LAO layers 
the calculated gap between the VBM of the LAO surface
layer and the STO CBM is 1.1 eV, which is too large to be closed
by water molecule adsorption because the VBM up-shift due to water adsorption
is only about 0.6 eV. In the case of four LAO layers, the interface
is already metallic and has a carrier density of 
$2\times10^{13}$ cm$^{-2}$ as reported by Thiel \textit{et
al.}\cite{thiel2006}, which may smooth over the carrier density change due to
the different adsorption configurations.

Noticeably, besides the aforementioned mechanism of conductance change at the
LAO/STO interface,  our results also imply another scheme to realize a
sharper switch between insulating and metal states at the LAO/STO
interface with three LAO layers by adsorbing and desorbing water
molecules at the LAO surface. In this scheme the insulating state
corresponds to the case of bare surface and the metallic state
corresponds to an adsorption configuration of 1 ML of water
molecules adsorbed at the surface. The insulator-metal transition
can be realized by exposing the bare surface to a water vapor and
the reverse transition can be realized by desorbing the water
molecules from the surface.

\section{Summary}
In summary, we have studied electronic and structural characteristics of
water molecule adsorption on the LAO surface of the $n$-type
LAO/STO interface by using first-principles electronic structure calculations.
We found that in an ambient atmosphere at room
temperature the adsorption configuration of 1 ML of water molecules including
3/4 ML of dissociated molecules is most stable and the configuration of 1 ML of
dissociated water molecules  is metastable. For the
LAO/STO interface with three LAO layers, dissociation of no less
than 0.5 ML of adsorbed water molecules results in a metallic interface,  and
the carrier density increases with
increasing the coverage of dissociated water molecules. 
A structural transition between the ground-state and a metastable
water adsorption configurations can induce a  carrier density change at the
LAO/STO interface. Our findings provide new insights into creating and
modulating electron carriers at the LAO/STO interface,
which hopefully prompt the application of oxide interface electronics. 
In addition, our results also suggest that water molecule adsorption from the
ambient atmosphere should be paid particular attention for experimentally
studying electronic properties of metal oxides, especially for the metal oxides
with a small band gap it may dramatically change some fundamental 
electronic properties of the materials, for instance, insulator-metal
transition. 

\section{acknowledgement}
This work was supported by the National Research Foundation of Korea
through the ARP (R17-2008-033-01000-0).  
We acknowledge the support from KISTI under the Supercomputing Application
Support Program.


%


\begin{thebibliography}{10}%
\makeatletter
\providecommand \@ifxundefined [1]{%
 \ifx #1\undefined \expandafter \@firstoftwo
 \else \expandafter \@secondoftwo
\fi
}%
\providecommand \@ifnum [1]{%
 \ifnum #1\expandafter \@firstoftwo
 \else \expandafter \@secondoftwo
\fi
}%
\providecommand \enquote [1]{``#1''}%
\providecommand \bibnamefont  [1]{#1}%
\providecommand \bibfnamefont [1]{#1}%
\providecommand \citenamefont [1]{#1}%
\providecommand\href[0]{\@sanitize\@href}%
\providecommand\@href[1]{\endgroup\@@startlink{#1}\endgroup\@@href}%
\providecommand\@@href[1]{#1\@@endlink}%
\providecommand \@sanitize [0]{\begingroup\catcode`\&12\catcode`\#12\relax}%
\@ifxundefined \pdfoutput {\@firstoftwo}{%
 \@ifnum{\z@=\pdfoutput}{\@firstoftwo}{\@secondoftwo}%
}{%
 \providecommand\@@startlink[1]{\leavevmode}%
 \providecommand\@@endlink[0]{}%
}{%
 \providecommand\@@startlink[1]{%
  \leavevmode
  \pdfstartlink
   attr{/Border[0 0 1 ]/H/I/C[0 1 1]}%
   user{/Subtype/Link/A<</Type/Action/S/URI/URI(#1)>>}%
  \relax
 }%
 \providecommand\@@endlink[0]{\pdfendlink}%
}%
\providecommand \url  [0]{\begingroup\@sanitize \@url }%
\providecommand \@url [1]{\endgroup\@href {#1}{\urlprefix}}%
\providecommand \urlprefix [0]{URL }%
\providecommand \Eprint[0]{\href }%
\@ifxundefined \urlstyle {%
  \providecommand \doi [1]{doi:\discretionary{}{}{}#1}%
}{%
  \providecommand \doi [0]{doi:\discretionary{}{}{}\begingroup
  \urlstyle{rm}\Url }%
}%
\providecommand \doibase [0]{http://dx.doi.org/}%
\providecommand \Doi[1]{\href{\doibase#1}}%
\providecommand \bibAnnote [3]{%
  \BibitemShut{#1}%
  \begin{quotation}\noindent
    \textsc{Key:}\ #2\\\textsc{Annotation:}\ #3%
  \end{quotation}%
}%
\providecommand \bibAnnoteFile [2]{%
  \IfFileExists{#2}{\bibAnnote {#1} {#2} {\input{#2}}}{}%
}%
\providecommand \typeout [0]{\immediate \write \m@ne }%
\providecommand \selectlanguage [0]{\@gobble}%
\providecommand \bibinfo [0]{\@secondoftwo}%
\providecommand \bibfield [0]{\@secondoftwo}%
\providecommand \translation [1]{[#1]}%
\providecommand \BibitemOpen[0]{}%
\providecommand \bibitemStop [0]{}%
\providecommand \bibitemNoStop [0]{.\EOS\space}%
\providecommand \EOS [0]{\spacefactor3000\relax}%
\providecommand \BibitemShut [1]{\csname bibitem#1\endcsname}%
\bibitem{ohtomo2004}%
  \BibitemOpen
  \bibfield{author}{%
  \bibinfo {author} {\bibfnamefont{A.}~\bibnamefont{Ohtomo}}\ and\ \bibinfo
  {author} {\bibfnamefont{H.~Y.}\ \bibnamefont{Hwang}},\ }%
  \bibfield{journal}{%
  \bibinfo {journal} {Nature}\ }%
  \textbf{\bibinfo {volume} {427}},\ \bibinfo {pages} {423} (\bibinfo {year}
  {2004})%
  \bibAnnoteFile{NoStop}{ohtomo2004}%
\bibitem{brinkman2007}%
  \BibitemOpen
  \bibfield{author}{%
  \bibinfo {author} {\bibfnamefont{A.}~\bibnamefont{Brinkman}}, \bibinfo
  {author} {\bibfnamefont{M.}~\bibnamefont{Huijben}}, \bibinfo {author}
  {\bibfnamefont{M.}~\bibnamefont{Van~Zalk}}, \bibinfo {author}
  {\bibfnamefont{J.}~\bibnamefont{Huijben}}, \bibinfo {author}
  {\bibfnamefont{U.}~\bibnamefont{Zeitler}}, \bibinfo {author}
  {\bibfnamefont{J.~C.}\ \bibnamefont{Maan}}, \bibinfo {author}
  {\bibfnamefont{W.~G.}\ \bibnamefont{Van~der Wiel}}, \bibinfo {author}
  {\bibfnamefont{G.}~\bibnamefont{Rijnders}}, \bibinfo {author}
  {\bibfnamefont{D.~H.~A.}\ \bibnamefont{Blank}},\ and\ \bibinfo {author}
  {\bibfnamefont{H.}~\bibnamefont{Hilgenkamp}},\ }%
  \bibfield{journal}{%
  \bibinfo {journal} {Nature Mater.}\ }%
  \textbf{\bibinfo {volume} {6}},\ \bibinfo {pages} {493} (\bibinfo {year}
  {2007})%
  \bibAnnoteFile{NoStop}{brinkman2007}%
\bibitem{reyren2007}%
  \BibitemOpen
  \bibfield{author}{%
  \bibinfo {author} {\bibfnamefont{N.}~\bibnamefont{Reyren}}, \bibinfo {author}
  {\bibfnamefont{S.}~\bibnamefont{Thiel}}, \bibinfo {author}
  {\bibfnamefont{A.~D.}\ \bibnamefont{Caviglia}}, \bibinfo {author}
  {\bibfnamefont{L.}~\bibnamefont{Fitting~Kourkoutis}}, \bibinfo {author}
  {\bibfnamefont{G.}~\bibnamefont{Hammerl}}, \bibinfo {author}
  {\bibfnamefont{C.}~\bibnamefont{Richter}}, \bibinfo {author}
  {\bibfnamefont{C.~W.}\ \bibnamefont{Schneider}}, \bibinfo {author}
  {\bibfnamefont{T.}~\bibnamefont{Kopp}}, \bibinfo {author}
  {\bibfnamefont{A.-S.}\ \bibnamefont{R\"uetschi}}, \bibinfo {author}
  {\bibfnamefont{D.}~\bibnamefont{Jaccard}}, \bibinfo {author}
  {\bibfnamefont{M.}~\bibnamefont{Gabay}}, \bibinfo {author}
  {\bibfnamefont{D.~A.}\ \bibnamefont{Muller}}, \bibinfo {author}
  {\bibfnamefont{J.-M.}\ \bibnamefont{Triscone}},\ and\ \bibinfo {author}
  {\bibfnamefont{J.}~\bibnamefont{Mannhart}},\ }%
  \bibfield{journal}{%
  \bibinfo {journal} {Science}\ }%
  \textbf{\bibinfo {volume} {317}},\ \bibinfo {pages} {1196} (\bibinfo {year}
  {2007})%
  \bibAnnoteFile{NoStop}{reyren2007}%
\bibitem{nakagawa2006}%
  \BibitemOpen
  \bibfield{author}{%
  \bibinfo {author} {\bibfnamefont{N.}~\bibnamefont{Nakagawa}}, \bibinfo
  {author} {\bibfnamefont{H.~Y.}\ \bibnamefont{Hwang}},\ and\ \bibinfo {author}
  {\bibfnamefont{D.~A.}\ \bibnamefont{Muller}},\ }%
  \bibfield{journal}{%
  \bibinfo {journal} {Nature Mater.}\ }%
  \textbf{\bibinfo {volume} {5}},\ \bibinfo {pages} {204} (\bibinfo {year}
  {2006})%
  \bibAnnoteFile{NoStop}{nakagawa2006}%
\bibitem{thiel2006}%
  \BibitemOpen
  \bibfield{author}{%
  \bibinfo {author} {\bibfnamefont{S.}~\bibnamefont{Thiel}}, \bibinfo {author}
  {\bibfnamefont{G.}~\bibnamefont{Hammerl}}, \bibinfo {author}
  {\bibfnamefont{A.}~\bibnamefont{Schmehl}}, \bibinfo {author}
  {\bibfnamefont{C.~W.}\ \bibnamefont{Schneider}},\ and\ \bibinfo {author}
  {\bibfnamefont{J.}~\bibnamefont{Mannhart}},\ }%
  \bibfield{journal}{%
  \bibinfo {journal} {Science}\ }%
  \textbf{\bibinfo {volume} {313}},\ \bibinfo {pages} {1942} (\bibinfo {year}
  {2006})%
  \bibAnnoteFile{NoStop}{thiel2006}%
\bibitem{pentcheva2009}%
  \BibitemOpen
  \bibfield{author}{%
  \bibinfo {author} {\bibfnamefont{R.}~\bibnamefont{Pentcheva}}\ and\ \bibinfo
  {author} {\bibfnamefont{W.~E.}\ \bibnamefont{Pickett}},\ }%
  \bibfield{journal}{%
  \bibinfo {journal} {Phys. Rev. Lett.}\ }%
  \textbf{\bibinfo {volume} {102}},\ \bibinfo {pages} {107602} (\bibinfo {year}
  {2009})%
  \bibAnnoteFile{NoStop}{pentcheva2009}%
\bibitem{li2010}%
  \BibitemOpen
  \bibfield{author}{%
  \bibinfo {author} {\bibfnamefont{Y.}~\bibnamefont{Li}}\ and\ \bibinfo
  {author} {\bibfnamefont{J.}~\bibnamefont{Yu}},\ }%
  \bibfield{journal}{%
  \bibinfo {journal} {J. Appl. Phys.}\ }%
  \textbf{\bibinfo {volume} {108}},\ \bibinfo {pages} {013701} (\bibinfo {year}
  {2010})%
  \bibAnnoteFile{NoStop}{li2010}%
\bibitem{herranz2007}%
  \BibitemOpen
  \bibfield{author}{%
  \bibinfo {author} {\bibfnamefont{G.}~\bibnamefont{Herranz}}, \bibinfo
  {author} {\bibfnamefont{M.}~\bibnamefont{Basleti\ifmmode~\acute{c}\else
  \'{c}\fi{}}}, \bibinfo {author} {\bibfnamefont{M.}~\bibnamefont{Bibes}},
  \bibinfo {author} {\bibfnamefont{C.}~\bibnamefont{Carr\'et\'ero}}, \bibinfo
  {author} {\bibfnamefont{E.}~\bibnamefont{Tafra}}, \bibinfo {author}
  {\bibfnamefont{E.}~\bibnamefont{Jacquet}}, \bibinfo {author}
  {\bibfnamefont{K.}~\bibnamefont{Bouzehouane}}, \bibinfo {author}
  {\bibfnamefont{C.}~\bibnamefont{Deranlot}}, \bibinfo {author}
  {\bibfnamefont{A.}~\bibnamefont{Hamzi\ifmmode~\acute{c}\else \'{c}\fi{}}},
  \bibinfo {author} {\bibfnamefont{J.-M.}\ \bibnamefont{Broto}}, \bibinfo
  {author} {\bibfnamefont{A.}~\bibnamefont{Barth\'el\'emy}},\ and\ \bibinfo
  {author} {\bibfnamefont{A.}~\bibnamefont{Fert}},\ }%
  \bibfield{journal}{%
  \bibinfo {journal} {Phys. Rev. Lett.}\ }%
  \textbf{\bibinfo {volume} {98}},\ \bibinfo {pages} {216803} (\bibinfo {year}
  {2007})%
  \bibAnnoteFile{NoStop}{herranz2007}%
\bibitem{kalabukhov2007}%
  \BibitemOpen
  \bibfield{author}{%
  \bibinfo {author} {\bibfnamefont{A.}~\bibnamefont{Kalabukhov}}, \bibinfo
  {author} {\bibfnamefont{R.}~\bibnamefont{Gunnarsson}}, \bibinfo {author}
  {\bibfnamefont{J.}~\bibnamefont{B\"orjesson}}, \bibinfo {author}
  {\bibfnamefont{E.}~\bibnamefont{Olsson}}, \bibinfo {author}
  {\bibfnamefont{T.}~\bibnamefont{Claeson}},\ and\ \bibinfo {author}
  {\bibfnamefont{D.}~\bibnamefont{Winkler}},\ }%
  \bibfield{journal}{%
  \bibinfo {journal} {Phys. Rev. B}\ }%
  \textbf{\bibinfo {volume} {75}},\ \bibinfo {pages} {121404} (\bibinfo {year}
  {2007})%
  \bibAnnoteFile{NoStop}{kalabukhov2007}%
\bibitem{li2011}%
  \BibitemOpen
  \bibfield{author}{%
  \bibinfo {author} {\bibfnamefont{Y.}~\bibnamefont{Li}}, \bibinfo {author}
  {\bibfnamefont{S.~N.}\ \bibnamefont{Phattalung}}, \bibinfo {author}
  {\bibfnamefont{S.}~\bibnamefont{Limpijumnong}}, \bibinfo {author}
  {\bibfnamefont{J.}~\bibnamefont{Kim}},\ and\ \bibinfo {author}
  {\bibfnamefont{J.}~\bibnamefont{Yu}},\ }%
  \bibfield{journal}{%
  \bibinfo {journal} {Phys. Rev. B}\ }%
  \textbf{\bibinfo {volume} {84}},\ \bibinfo {pages} {245307} (\bibinfo {month}
  {Dec}\ \bibinfo {year} {2011})%
  \bibAnnoteFile{NoStop}{li2011}%
\bibitem{cen2008}%
  \BibitemOpen
  \bibfield{author}{%
  \bibinfo {author} {\bibfnamefont{C.}~\bibnamefont{Cen}}, \bibinfo {author}
  {\bibfnamefont{S.}~\bibnamefont{Thiel}}, \bibinfo {author}
  {\bibfnamefont{G.}~\bibnamefont{Hammerl}}, \bibinfo {author}
  {\bibfnamefont{C.~W.}\ \bibnamefont{Schneider}}, \bibinfo {author}
  {\bibfnamefont{K.~E.}\ \bibnamefont{Andersen}}, \bibinfo {author}
  {\bibfnamefont{C.~S.}\ \bibnamefont{Hellberg}}, \bibinfo {author}
  {\bibfnamefont{J.}~\bibnamefont{Mannhart}},\ and\ \bibinfo {author}
  {\bibfnamefont{J.}~\bibnamefont{Levy}},\ }%
  \bibfield{journal}{%
  \bibinfo {journal} {Nature Mater.}\ }%
  \textbf{\bibinfo {volume} {7}},\ \bibinfo {pages} {298} (\bibinfo {year}
  {2008})%
  \bibAnnoteFile{NoStop}{cen2008}%
\bibitem{cen2009}%
  \BibitemOpen
  \bibfield{author}{%
  \bibinfo {author} {\bibfnamefont{C.}~\bibnamefont{Cen}}, \bibinfo {author}
  {\bibfnamefont{S.}~\bibnamefont{Thiel}}, \bibinfo {author}
  {\bibfnamefont{J.}~\bibnamefont{Mannhart}},\ and\ \bibinfo {author}
  {\bibfnamefont{J.}~\bibnamefont{Levy}},\ }%
  \bibfield{journal}{%
  \bibinfo {journal} {Science}\ }%
  \textbf{\bibinfo {volume} {323}},\ \bibinfo {pages} {1026} (\bibinfo {year}
  {2009})%
  \bibAnnoteFile{NoStop}{cen2009}%
\bibitem{bi2010}%
  \BibitemOpen
  \bibfield{author}{%
  \bibinfo {author} {\bibfnamefont{F.}~\bibnamefont{Bi}}, \bibinfo {author}
  {\bibfnamefont{D.~F.}\ \bibnamefont{Bogorin}}, \bibinfo {author}
  {\bibfnamefont{C.}~\bibnamefont{Cen}}, \bibinfo {author}
  {\bibfnamefont{C.~W.}\ \bibnamefont{Bark}}, \bibinfo {author}
  {\bibfnamefont{J.~W.}\ \bibnamefont{Park}}, \bibinfo {author}
  {\bibfnamefont{C.~B.}\ \bibnamefont{Eom}},\ and\ \bibinfo {author}
  {\bibfnamefont{J.}~\bibnamefont{Levy}},\ }%
  \bibfield{journal}{%
  \bibinfo {journal} {Applied Physics Letters}\ }%
  \textbf{\bibinfo {volume} {97}},\ \bibinfo {pages} {173110} (\bibinfo {year}
  {2010})%
  \bibAnnoteFile{NoStop}{bi2010}%
\bibitem{xie2011}%
  \BibitemOpen
  \bibfield{author}{%
  \bibinfo {author} {\bibfnamefont{Y.}~\bibnamefont{Xie}}, \bibinfo {author}
  {\bibfnamefont{Y.}~\bibnamefont{Hikita}}, \bibinfo {author}
  {\bibfnamefont{C.}~\bibnamefont{Bell}},\ and\ \bibinfo {author}
  {\bibfnamefont{H.~Y.}\ \bibnamefont{Hwang}},\ }%
  \bibfield{journal}{%
  \bibinfo {journal} {Nature Communications}\ }%
  \textbf{\bibinfo {volume} {2}},\ \bibinfo {pages} {494} (\bibinfo {year}
  {2011})%
  \bibAnnoteFile{NoStop}{xie2011}%
\bibitem{son2010}%
  \BibitemOpen
  \bibfield{author}{%
  \bibinfo {author} {\bibfnamefont{W.}~\bibnamefont{Son}}, \bibinfo {author}
  {\bibfnamefont{E.}~\bibnamefont{Cho}}, \bibinfo {author}
  {\bibfnamefont{J.}~\bibnamefont{Lee}},\ and\ \bibinfo {author}
  {\bibfnamefont{S.}~\bibnamefont{Han}},\ }%
  \bibfield{journal}{%
  \bibinfo {journal} {Journal of Physics: Condensed Matter}\ }%
  \textbf{\bibinfo {volume} {22}},\ \bibinfo {pages} {315501} (\bibinfo {year}
  {2010})%
  \bibAnnoteFile{NoStop}{son2010}%
\bibitem{thiel1987}%
  \BibitemOpen
  \bibfield{author}{%
  \bibinfo {author} {\bibfnamefont{P.~A.}\ \bibnamefont{Thiel}}\ and\ \bibinfo
  {author} {\bibfnamefont{T.~E.}\ \bibnamefont{Madey}},\ }%
  \bibfield{journal}{%
  \bibinfo {journal} {Surface Science Reports}\ }%
  \textbf{\bibinfo {volume} {7}},\ \bibinfo {pages} {211} (\bibinfo {year}
  {1987})%
  \bibAnnoteFile{NoStop}{thiel1987}%
\bibitem{hass1998}%
  \BibitemOpen
  \bibfield{author}{%
  \bibinfo {author} {\bibfnamefont{K.~C.}\ \bibnamefont{Hass}}, \bibinfo
  {author} {\bibfnamefont{W.~F.}\ \bibnamefont{Schneider}}, \bibinfo {author}
  {\bibfnamefont{A.}~\bibnamefont{Curioni}},\ and\ \bibinfo {author}
  {\bibfnamefont{W.}~\bibnamefont{Andreoni}},\ }%
  \bibfield{journal}{%
  \bibinfo {journal} {Science}\ }%
  \textbf{\bibinfo {volume} {282}},\ \bibinfo {pages} {265} (\bibinfo {year}
  {1998})%
  \bibAnnoteFile{NoStop}{hass1998}%
\bibitem{shapovalov2000}%
  \BibitemOpen
  \bibfield{author}{%
  \bibinfo {author} {\bibfnamefont{V.}~\bibnamefont{Shapovalov}}\ and\ \bibinfo
  {author} {\bibfnamefont{T.~N.}\ \bibnamefont{Truong}},\ }%
  \bibfield{journal}{%
  \bibinfo {journal} {The Journal of Physical Chemistry B}\ }%
  \textbf{\bibinfo {volume} {104}},\ \bibinfo {pages} {9859} (\bibinfo {year}
  {2000})%
  \bibAnnoteFile{NoStop}{shapovalov2000}%
\bibitem{henderson2002}%
  \BibitemOpen
  \bibfield{author}{%
  \bibinfo {author} {\bibfnamefont{M.~A.}\ \bibnamefont{Henderson}},\ }%
  \bibfield{journal}{%
  \bibinfo {journal} {Surface Science Reports}\ }%
  \textbf{\bibinfo {volume} {46}},\ \bibinfo {pages} {1} (\bibinfo {year}
  {2002})%
  \bibAnnoteFile{NoStop}{henderson2002}%
\bibitem{ionescu2002}%
  \BibitemOpen
  \bibfield{author}{%
  \bibinfo {author} {\bibfnamefont{A.}~\bibnamefont{Ionescu}}, \bibinfo
  {author} {\bibfnamefont{A.}~\bibnamefont{Allouche}}, \bibinfo {author}
  {\bibfnamefont{J.~P.}\ \bibnamefont{Aycard}}, \bibinfo {author}
  {\bibfnamefont{M.}~\bibnamefont{Rajzmann}},\ and\ \bibinfo {author}
  {\bibfnamefont{F.}~\bibnamefont{Hutschka}},\ }%
  \bibfield{journal}{%
  \bibinfo {journal} {The Journal of Physical Chemistry B}\ }%
  \textbf{\bibinfo {volume} {106}},\ \bibinfo {pages} {9359} (\bibinfo {year}
  {2002})%
  \bibAnnoteFile{NoStop}{ionescu2002}%
\bibitem{evarestov2007}%
  \BibitemOpen
  \bibfield{author}{%
  \bibinfo {author} {\bibfnamefont{R.~A.}\ \bibnamefont{Evarestov}}, \bibinfo
  {author} {\bibfnamefont{A.~V.}\ \bibnamefont{Bandura}},\ and\ \bibinfo
  {author} {\bibfnamefont{V.~E.}\ \bibnamefont{Alexandrov}},\ }%
  \bibfield{journal}{%
  \bibinfo {journal} {Surface science}\ }%
  \textbf{\bibinfo {volume} {601}},\ \bibinfo {pages} {1844} (\bibinfo {year}
  {2007})%
  \bibAnnoteFile{NoStop}{evarestov2007}%
\bibitem{ranea2008}%
  \BibitemOpen
  \bibfield{author}{%
  \bibinfo {author} {\bibfnamefont{V.~A.}\ \bibnamefont{Ranea}}, \bibinfo
  {author} {\bibfnamefont{W.~F.}\ \bibnamefont{Schneider}},\ and\ \bibinfo
  {author} {\bibfnamefont{I.}~\bibnamefont{Carmichael}},\ }%
  \bibfield{journal}{%
  \bibinfo {journal} {Surface Science}\ }%
  \textbf{\bibinfo {volume} {602}},\ \bibinfo {pages} {268} (\bibinfo {year}
  {2008})%
  \bibAnnoteFile{NoStop}{ranea2008}%
\bibitem{baniecki2009}%
  \BibitemOpen
  \bibfield{author}{%
  \bibinfo {author} {\bibfnamefont{J.~D.}\ \bibnamefont{Baniecki}}, \bibinfo
  {author} {\bibfnamefont{M.}~\bibnamefont{Ishii}}, \bibinfo {author}
  {\bibfnamefont{K.}~\bibnamefont{Kurihara}}, \bibinfo {author}
  {\bibfnamefont{K.}~\bibnamefont{Yamanaka}}, \bibinfo {author}
  {\bibfnamefont{T.}~\bibnamefont{Yano}}, \bibinfo {author}
  {\bibfnamefont{K.}~\bibnamefont{Shinozaki}}, \bibinfo {author}
  {\bibfnamefont{T.}~\bibnamefont{Imada}},\ and\ \bibinfo {author}
  {\bibfnamefont{Y.}~\bibnamefont{Kobayashi}},\ }%
  \bibfield{journal}{%
  \bibinfo {journal} {Journal of Applied Physics}\ }%
  \textbf{\bibinfo {volume} {106}},\ \bibinfo {pages} {054109} (\bibinfo {year}
  {2009})%
  \bibAnnoteFile{NoStop}{baniecki2009}%
\bibitem{guhl2010}%
  \BibitemOpen
  \bibfield{author}{%
  \bibinfo {author} {\bibfnamefont{H.}~\bibnamefont{Guhl}}, \bibinfo {author}
  {\bibfnamefont{W.}~\bibnamefont{Miller}},\ and\ \bibinfo {author}
  {\bibfnamefont{K.}~\bibnamefont{Reuter}},\ }%
  \bibfield{journal}{%
  \bibinfo {journal} {Physical Review B}\ }%
  \textbf{\bibinfo {volume} {81}},\ \bibinfo {pages} {155455} (\bibinfo {year}
  {2010})%
  \bibAnnoteFile{NoStop}{guhl2010}%
\bibitem{emsley1980}%
  \BibitemOpen
  \bibfield{author}{%
  \bibinfo {author} {\bibfnamefont{J.}~\bibnamefont{Emsley}},\ }%
  \bibfield{journal}{%
  \bibinfo {journal} {Chemical Society Reviews}\ }%
  \textbf{\bibinfo {volume} {9}},\ \bibinfo {pages} {91} (\bibinfo {year}
  {1980})%
  \bibAnnoteFile{NoStop}{emsley1980}%
\bibitem{rijnders2008}%
  \BibitemOpen
  \bibfield{author}{%
  \bibinfo {author} {\bibfnamefont{G.}~\bibnamefont{Rijnders}}\ and\ \bibinfo
  {author} {\bibfnamefont{D.~H.~A.}\ \bibnamefont{Blank}},\ }%
  \bibfield{journal}{%
  \bibinfo {journal} {Nature Mater.}\ }%
  \textbf{\bibinfo {volume} {7}},\ \bibinfo {pages} {270} (\bibinfo {year}
  {2008})%
  \bibAnnoteFile{NoStop}{rijnders2008}%
\bibitem{kresse1996}%
  \BibitemOpen
  \bibfield{author}{%
  \bibinfo {author} {\bibfnamefont{G.}~\bibnamefont{Kresse}}\ and\ \bibinfo
  {author} {\bibfnamefont{J.}~\bibnamefont{Furthm\"uller}},\ }%
  \bibfield{journal}{%
  \bibinfo {journal} {Phys. Rev. B}\ }%
  \textbf{\bibinfo {volume} {54}},\ \bibinfo {pages} {11169} (\bibinfo {month}
  {Oct}\ \bibinfo {year} {1996})%
  \bibAnnoteFile{NoStop}{kresse1996}%
\bibitem{wang1991}%
  \BibitemOpen
  \bibfield{author}{%
  \bibinfo {author} {\bibfnamefont{Y.}~\bibnamefont{Wang}}\ and\ \bibinfo
  {author} {\bibfnamefont{J.~P.}\ \bibnamefont{Perdew}},\ }%
  \bibfield{journal}{%
  \bibinfo {journal} {Phys. Rev. B}\ }%
  \textbf{\bibinfo {volume} {44}},\ \bibinfo {pages} {13298} (\bibinfo {year}
  {1991})%
  \bibAnnoteFile{NoStop}{wang1991}%
\bibitem{blochl1994}%
  \BibitemOpen
  \bibfield{author}{%
  \bibinfo {author} {\bibfnamefont{P.~E.}\ \bibnamefont{Bl\"ochl}},\ }%
  \bibfield{journal}{%
  \bibinfo {journal} {Phys. Rev. B}\ }%
  \textbf{\bibinfo {volume} {50}},\ \bibinfo {pages} {17953} (\bibinfo {year}
  {1994})%
  \bibAnnoteFile{NoStop}{blochl1994}%
\bibitem{kresse1999}%
  \BibitemOpen
  \bibfield{author}{%
  \bibinfo {author} {\bibfnamefont{G.}~\bibnamefont{Kresse}}\ and\ \bibinfo
  {author} {\bibfnamefont{D.}~\bibnamefont{Joubert}},\ }%
  \bibfield{journal}{%
  \bibinfo {journal} {Phys. Rev. B}\ }%
  \textbf{\bibinfo {volume} {59}},\ \bibinfo {pages} {1758} (\bibinfo {year}
  {1999})%
  \bibAnnoteFile{NoStop}{kresse1999}%
\bibitem{makov1995}%
  \BibitemOpen
  \bibfield{author}{%
  \bibinfo {author} {\bibfnamefont{G.}~\bibnamefont{Makov}}\ and\ \bibinfo
  {author} {\bibfnamefont{M.~C.}\ \bibnamefont{Payne}},\ }%
  \bibfield{journal}{%
  \bibinfo {journal} {Phys. Rev. B}\ }%
  \textbf{\bibinfo {volume} {51}},\ \bibinfo {pages} {4014} (\bibinfo {year}
  {1995})%
  \bibAnnoteFile{NoStop}{makov1995}%
\bibitem{stull1971}%
  \BibitemOpen
  \emph{\bibinfo {title} {JANAF Thermochemical Tables}},\ \bibinfo {edition}
  {2nd}\ ed.,\ edited by\ \bibinfo {editor} {\bibfnamefont{D.~R.}\
  \bibnamefont{Stull}}\ and\ \bibinfo {editor}
  {\bibfnamefont{H.}~\bibnamefont{Prophet}}\ (\bibinfo {publisher} {U.S.
  National Bureau of Standards},\ \bibinfo {address} {Washington, D. C.},\
  \bibinfo {year} {1971})%
  \bibAnnoteFile{NoStop}{stull1971}%
\bibitem{shin2010}%
  \BibitemOpen
  \bibfield{author}{%
  \bibinfo {author} {\bibfnamefont{H.~J.}\ \bibnamefont{Shin}}, \bibinfo
  {author} {\bibfnamefont{J.}~\bibnamefont{Jung}}, \bibinfo {author}
  {\bibfnamefont{K.}~\bibnamefont{Motobayashi}}, \bibinfo {author}
  {\bibfnamefont{S.}~\bibnamefont{Yanagisawa}}, \bibinfo {author}
  {\bibfnamefont{Y.}~\bibnamefont{Morikawa}}, \bibinfo {author}
  {\bibfnamefont{Y.}~\bibnamefont{Kim}},\ and\ \bibinfo {author}
  {\bibfnamefont{M.}~\bibnamefont{Kawai}},\ }%
  \bibfield{journal}{%
  \bibinfo {journal} {Nature materials}\ }%
  \textbf{\bibinfo {volume} {9}},\ \bibinfo {pages} {442} (\bibinfo {year}
  {2010})%
  \bibAnnoteFile{NoStop}{shin2010}%
\end{thebibliography}
\end{document}